\documentclass{article}
\usepackage{arxiv}

\usepackage[utf8]{inputenc} 
\usepackage[T1]{fontenc}    
\usepackage{hyperref}       
\usepackage{url}            
\usepackage{booktabs}       
\usepackage{amsfonts}       
\usepackage{nicefrac}       
\usepackage{microtype}      
\usepackage{lipsum}         
\usepackage{graphicx}
\usepackage{natbib}
\usepackage{hyperref}
\usepackage{xcolor}
\usepackage{makecell}

\usepackage{doi}

\usepackage{amsmath,amssymb}
\usepackage{cleveref}       

\usepackage{appendix} 

\title{A fine-tuning workflow for automatic first-break picking with deep learning}

\newcommand{\authorHref}[3][black]{\href{#2}{\color{#1}{#3}}}

\author{
\authorHref[black]{https://orcid.org/0000-0002-0417-1259}{\includegraphics[scale=0.06]{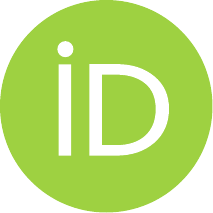}\hspace{1mm}Amir Mardan}\thanks{Corresponding author, \authorHref[black]{mardan.amir.h@gmail.com}{mardan.amir.h@gmail.com}} \\
	Geostack\\
	Polytechnique Montréal\\
	QC, Canada \\
	\texttt{mardan.amir.h@gmail.com}\\
	\And
	Martin Blouin \\
	Geostack\\
	Québec, QC, Canada \\
	\texttt{mblouin@geostack.ca}\\
	\And
	\authorHref{https://orcid.org/0000-0002-1849-3718}{\includegraphics[scale=0.06]{orcid.pdf}\hspace{1mm}Gabriel Fabien-Ouellet} \\
	Polytechnique Montréal\\
	Montréal, QC, Canada \\
	\texttt{gabriel.fabien-ouellet@polymtl.ca} \\
	\And
\authorHref{https://orcid.org/0000-0002-2042-2759}{\includegraphics[scale=0.06]{orcid.pdf}\hspace{1mm}Bernard Giroux} \\
	INRS-ETE\\
	Québec, QC, Canada \\
	\texttt{bernard.giroux@inrs.ca}\\
	\And
	Christophe Vergniault\\
	Électricité de France\\
	France \\
	\texttt{christophe.vergniault@edf.fr}\\
	\And
	Jeremy Gendreau \\
	Polytechnique Montréal\\
	Montréal, QC, Canada \\
	\texttt{jeremy.gendreau@polymtl.ca}
}

\renewcommand{\undertitle}{Fine-tuning workflow for automatic first-break picking}
\renewcommand{\shorttitle}{Fine-tuning workflow for automatic first-break picking}

\hypersetup{
colorlinks   = true,
linkcolor    = blue,
urlcolor = blue,
citecolor = blue,
pdftitle={Fine-tuning for automatic first breaking picking},
pdfsubject={q-bio.NC, q-bio.QM},
pdfauthor={Amir Mardan},
pdfkeywords={First-break picking workflow, Near-surface seismic, Deep learning, Segmentation, U-Net},
}

\graphicspath{{./figs}}

\begin{document}

\maketitle

\begin{abstract}
First-break picking is an essential step in seismic data processing. 
For reliable results, first arrivals should be picked by an expert. 
This is a time-consuming procedure and subjective to a certain degree, leading to different results for different operators. 
In this study, we have used a U-Net architecture with residual blocks to perform automatic first-break picking based on deep learning. 
Focusing on the effects of weight initialization on first-break picking, we conduct this research by using the weights of a pretrained network that is used for object detection on the ImageNet dataset. The efficiency of the proposed method is tested on two real datasets. 
For both datasets, we pick manually the first breaks for less than 10\% of the seismic shots. 
The pretrained network is fine-tuned on the picked shots and the rest of the shots are automatically picked by the neural network. 
It is shown that this strategy allows to reduce the size of the training set, requiring fine tuning with only a few picked shots per survey. 
Using random weights and more training epochs can lead to a lower training loss, but such a strategy leads to overfitting as the test error is higher than the one of the pretrained network. 
We also assess the possibility of using a general dataset by training a network with data from three different projects that are acquired with different equipment and at different locations. 
This study shows that if the general dataset is created carefully it can lead to more accurate first-break picking, otherwise the general dataset can decrease the accuracy. 
Focusing on near-surface geophysics, we perform traveltime tomography and compare the inverted velocity models based on different first-break picking methodologies. 
The results of the inversion show that the first breaks obtained by the pretrained network lead to a velocity model that is closer to the one obtained from the inversion of expert-picked first breaks. 
\end{abstract}

\keywords{First-break picking, Near-surface seismic, Deep learning,
Segmentation, Transfer learning}

\section*{Introduction}
In a seismic shot record, the first arrival is usually the direct wave from the source followed by refractions at the base of the weathering layer. 
The first break (FB) is the time taken by  a seismic wave to travel from a source to a geophone. 
First breaks are picked to estimate static corrections \citep{coppens1985first,marsden1993static}, near-surface velocity estimation \citep{azwin2013applying,Fabien_OuelletEtAl2014using}, muting parameters for processing reflected data, and microseismic event location \citep{dip2021microseismic, nasr2022python}.
In marine studies, first breaks can be studied to also estimate the location of hydrophones \cite[]{walia1999source} as well as modeling the water velocity variation, which is highly important for time-lapse seismic studies \citep{MardanEtAl2023wa_geophysics}.

\begin{figure}[ht]
\begin{center}\includegraphics[width=1\textwidth]{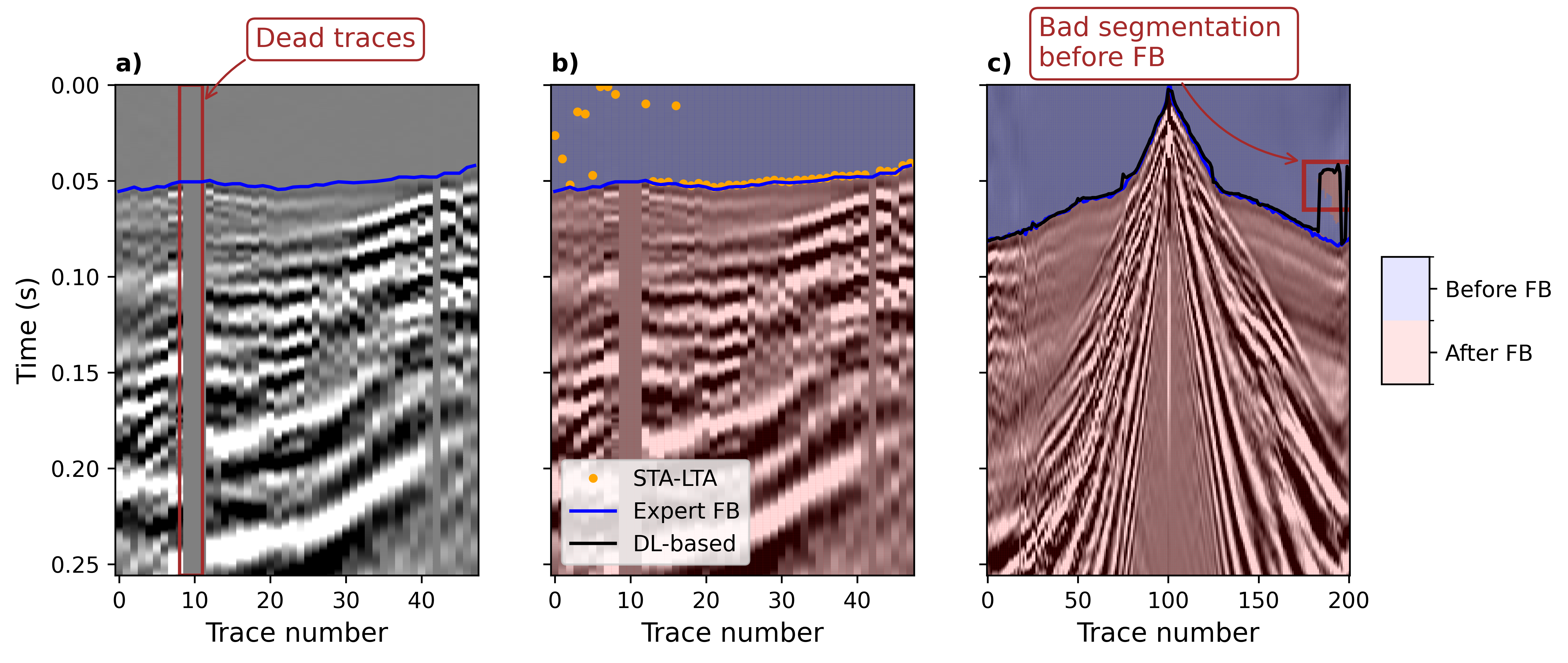}
\caption{An example of automatic first-break picking, its challenges, and seismic shot segmentation. (a) An example of seismic shot with corresponding FB picked by an expert (blue line). Red rectangle shows some dead traces whose corresponding geophones have not been active during the acquisition. (b) A shot gather with first breaks picked by an expert and STA-LTA method (orange dot). (c) In data segmentation a shot gather is divided into two segments as before FB (blue) and after FB (red) whose interface is the predicted FB as shown with black line. Bad segmentation is shown with red rectangle.
{\label{fig:challenges}}
}
\end{center}
\end{figure}

First breaks are conventionally picked manually.
Manual picking is a labor-intensive and time-consuming processing step which can take up to 30$\%$ of the processing time \cite[]{sabbione2010automatic}.
This is also a subjective task that leads to different results for different operators.
To mitigate this problem, numerous methods have been introduced with the aim of automating the process.
\cite{zwartjes2022first} classified these methods into three classes; cross-correlation based, energy-based, and neural-network (NN) based.
\cite{peraldi1972digital} proposed a method to automate the first-break picking based on the cross-correlation of adjacent traces in a seismic shot.
Finding the maximum cross-correlation between traces, they can estimate the time delay between first breaks in different traces.
This method has difficulty to detect first breaks efficiently in a seismic shot with bad or dead traces (Figure \ref{fig:challenges}a).
\cite{coppens1985first} proposed a method to automate first-break picking based on the energy ratio of the signal in two windows with different sizes. 
This method is called short-term-average over long-term average (STA-LTA) and its efficiency decreases in noisy data (Figure \ref{fig:challenges}b).
Although other methods such as modified Coppens' method \citep{sabbione2010automatic}, entropy method \citep{sabbione2010automatic}, and statistical methods \citep{hatherly1982computer} have been proposed to improve the accuracy of automatic first-break picking, these methods usually require an expert to select parameters such as window length and threshold among others, which increases human bias in this processing step.

Nowadays, numerous seismic data acquisitions are carried out every day around the globe using efficient techniques such as distributed-acoustic sensing (DAS) or landstreamers, with the consequence that  the volume of seismic data that must be processed is significantly increasing.
In this regard, artificial intelligence (AI) has been used to improve the seismic data processing and interpretation workflow \citep{zhao1988minimum, veezhinathan1990neural,roth1994neural,lim2005reservoir,leite20113d, mardan2017channel,fabien2020seismic,alali2022deep,mardan2024piann_eage}.
With the focus on automatic first-break picking, researchers started employing NN in the early 1990s \cite[]{veezhinathan1990neural, murat1992automated, mccormack1993first}.
With subsequent development in computational power and NN techniques, convolutional neural network (CNN) has been employed to improve the accuracy of automatic picking by taking the spatial coherency of data into account.
\cite{yuan2018seismic} employed a CNN to classify the data on a seismic shot into two groups: FB and non-FB.
They showed the potential of CNN for first-break picking. 
However, preparing the training data for this strategy is challenging and time consuming (if not more than manual first-break picking).
To mitigate this problem, first-break picking has been considered as a semantic segmentation problem. 
\cite{hu2019first} employed a U-Net \citep{ronneberger2015u} architecture to train a network for first-break picking based on segmenting a seismic section into FB and non-FB. 
They widened the FB segment to three pixels around picked first-break to increase the accuracy. 
Their workflow does not require picking non-FB for data preparation.
Nevertheless, it suffers from data imbalance which is a common challenge for image segmentation \citep{hossain2021dual}. 

As is shown in Figure \ref{fig:challenges}c, a seismic shot can be segmented into before and after FB. 
In this way, FB can be automatically picked as the interface between two segments.
Considering first-break picking as a segmentation problem, \cite{wu2019semiautomatic} and \cite{yuan2022segnet} employed SegNet \citep{badrinarayanan2017segnet} to automate the procedure.
\cite{wu2019semiautomatic} trained SegNet on traces of microseismic sections to predict first-arrivals. 
They assessed the efficiency of SegNet for detecting first arrivals in synthetic data with different signal-to-noise ratio (SNR) and in field data.
They showed the superiority of CNN-based method over STA-LTA method.
\cite{yuan2022segnet} employed a SegNet architecture to autopick first breaks on real seismic shots with sparsely distributed traces.
They showed that the CNN-based methods can give good performance for processing sparse data.

U-Net is another network architecture that has been used to pick first breaks accurately by performing segmentation.
 \cite{zhu2019phasenet} introduced PhaseNet by modifying U-Net to process 1D time-series data.
 For an accurate first-arrival picking of $P$- and $S$-waves, they picked the first arrivals manually in 700,920 traces that were used in the training process.
Focusing on earthquake data, \cite{zhu2019phasenet} showed that the trained PhaseNet was able to provide higher accuracy than STA-LTA.
\cite{maEtAl2020} used a U-Net to automate first-break picking in microseismic and borehole studies.
They investigated the effects of noise level in data and shape of the input data (1D or 2D seismic data) on the performance of the network.
\cite{zwartjes2022first} compared different architectures and training parameters aiming to find crucial parameters in CNN-based automatic first-break picking.
Comparing auto-encoder, U-Net, and U-Net with residual blocks, \cite{zwartjes2022first} showed that standard U-Net is computationally less demanding, while U-Net with residual blocks can lead to a better generalization and requires fewer epochs to reach a specific loss.

The aforementioned methods have different degree of accuracy in the presence of noise.
Effects of noise on results are more pronounced when receivers record strong noise before the arrival of direct and refracted waves (Figure \ref{fig:challenges}c).
Although DL has shown the highest accuracy to automate the first-break picking compared to cross-correlation and energy-based methods, it can be seen in the mentioned studies that the presence of noise before FBs remains challenging.
Performance of DL-based methods also relies drastically on the size of the dataset.
The motivation of this study is to increase the accuracy of automatic first-break picking while decreasing the required size of the training dataset.
This makes the DL-based FB picking applicable to near surface surveys, because training from scratch and building a training set with usual strategy is presently too costly for near surface studies.
To achieve this goal, we employ transfer learning \citep{bozinovski1976influence, goodfellow2016deep}.
Transfer learning is a machine learning technique where the knowledge that a model has obtained to perform a task can be used to improve the learning of a new task \citep{goodfellow2016deep}.
This technique has shown its potential to mitigate the challenges of using AI in geophysical problems \citep{cunha2020seismic, parkAndSacchi2020Automatic,simon2023hierarchical}.
In these studies, authors first train a network on a simple or synthetic dataset and then fine tune the parameters of the network with more complex or field data.

All previous studies on DL-based first-break picking use random initialization.
In this work, we investigate if the knowledge that a network has obtained from solving a non-geoscientific problem can be helpful for solving first-break picking.
In this regard, we design an architecture based on U-Net \citep{ronneberger2015u} with residual blocks \citep{he2016deep} as its encoder, to perform automatic first-break picking.
We show how the weights used for solving the ImageNet challenge \citep{deng2009imagenet} can increase the segmentation accuracy and lead to a more accurate first-break picking and reduce the size of the required training set to attain a fixed accuracy. 
The small size of the required dataset for training allows to fine tune the network on a per survey basis which makes our method applicable to most near surface surveys.  
We propose a workflow in which a small fraction (around 10$\%$) of the survey is picked by a human expert, then the NN is fine tuned and picks automatically the rest of the survey. 
In the current study, we analyze this strategy by applying the methodology to two refraction surveys used for near surface investigation.


This paper is organized as follows. First, we present the methodology including data preparation, architecture of the neural network, and loss function.
Next, we study the efficiency of the proposed method to pick FBs using two different datasets (with high and low SNR).
We then show how creating a general dataset can affect first-break picking by comparing the accuracy after training using a project-based dataset and using a general dataset. 
At the end, we assess the quality of the automatically picked first breaks by comparing velocity models obtained with travel time tomography.
 
\section*{Methodology}

\subsection*{Data preparation}
The key to have an efficient network that can predict with high accuracy is training with a suitable dataset.
A network should be trained with a large amount of data and the data should represent all possible scenarios that might exist in the real-world problem that is tackled.
Considering first-break picking as a segmentation problem, we need pairs of seismic shots and their corresponding FBs to train the network. 

\begin{figure}[t]
\begin{center}
\includegraphics[width=1.0\textwidth]{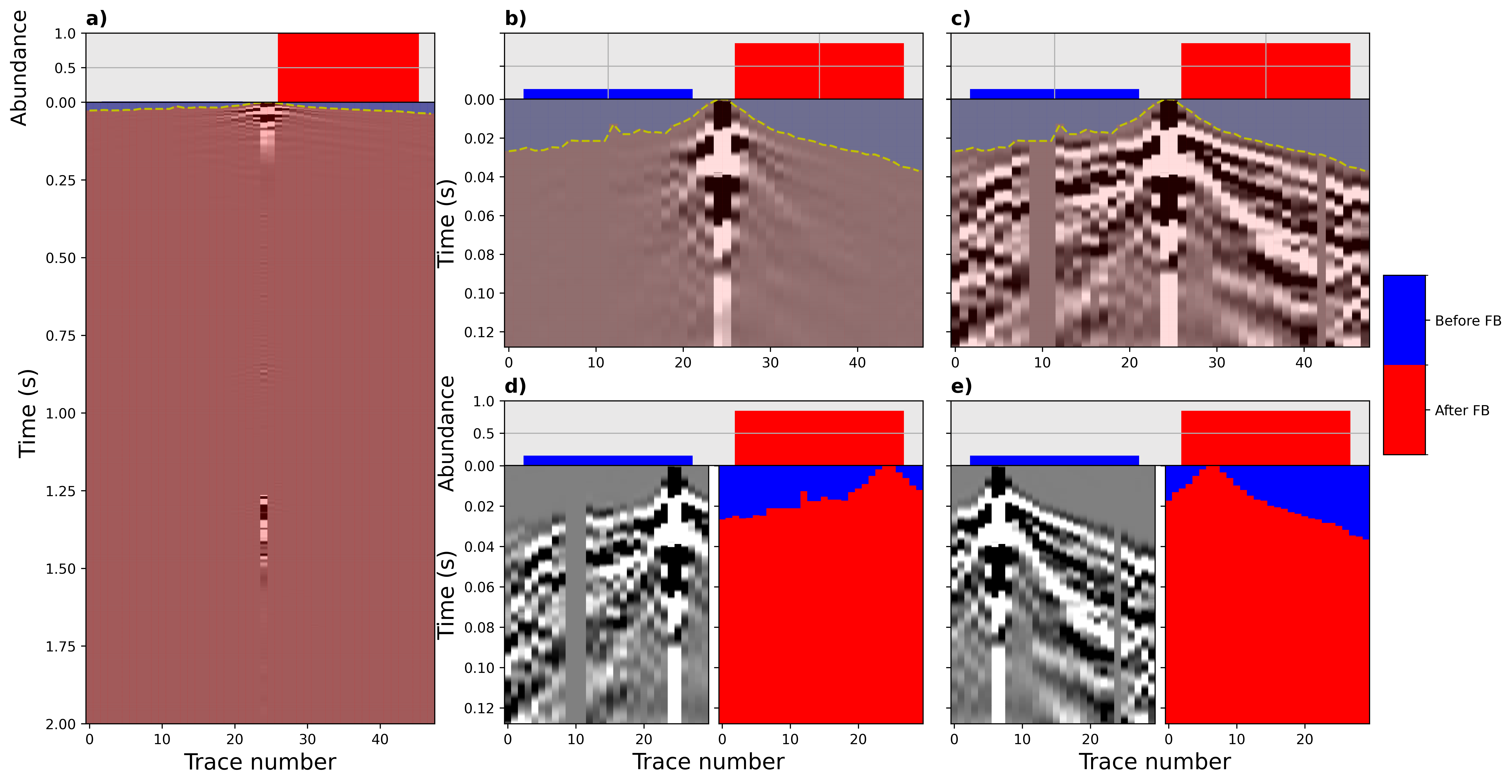}
\caption{The data preparation strategy employed for this study. (a) A raw seismic shot  is (b) cropped, (c) scaled, and (d-e) divided to subimages. The relative abundance of data in each segment is shown as bar plots above each seismic shot where blue indicates before FB and red demonstrates after FB. Yellow line shows the manually picked FB.
{\label{fig:data_preparing}}
}
\end{center}
\end{figure}

A common challenge in segmentation is class imbalance which reduces the accuracy of the network to predict the right segment for the minority-segment region \citep{hossain2021dual}.
Class imbalance means that the number of pixels in one segment is significantly higher than the number of pixels in other segments.
Figure \ref{fig:data_preparing}a shows a raw seismic shot (bottom) that is used in this study for training the network.
The relative abundance of data from each segment is shown in the top part of Figure \ref{fig:data_preparing}a.
As can be seen, almost all data samples belong to after FB ($99.0\%$).
Due to the fact that we are only interested in finding the first breaks, we can crop the seismic shot to improve the balance between segments (Figure \ref{fig:data_preparing}b).
This reduces the relative abundance of after FB to $84.7\%$.

In the next step, we scale each trace by its maximum amplitude (Figure \ref{fig:data_preparing}c) to facilitate learning in the training phase.
In the final step and based on number of traces in each seismic shot, shot gathers are divided into smaller sections with a specific overlap (Figure \ref{fig:data_preparing}d and \ref{fig:data_preparing}e). 
This data augmentation process can improve accuracy of the network by diversifying the training data.
At the end of processing, pairs of image-label are obtained from a raw seismic shot (Figure \ref{fig:data_preparing}a) as shown in Figure \ref{fig:data_preparing}d and \ref{fig:data_preparing}e.

\subsection*{Network architecture}
\begin{figure}[t]
\begin{center}
\includegraphics[width=1.0\textwidth]{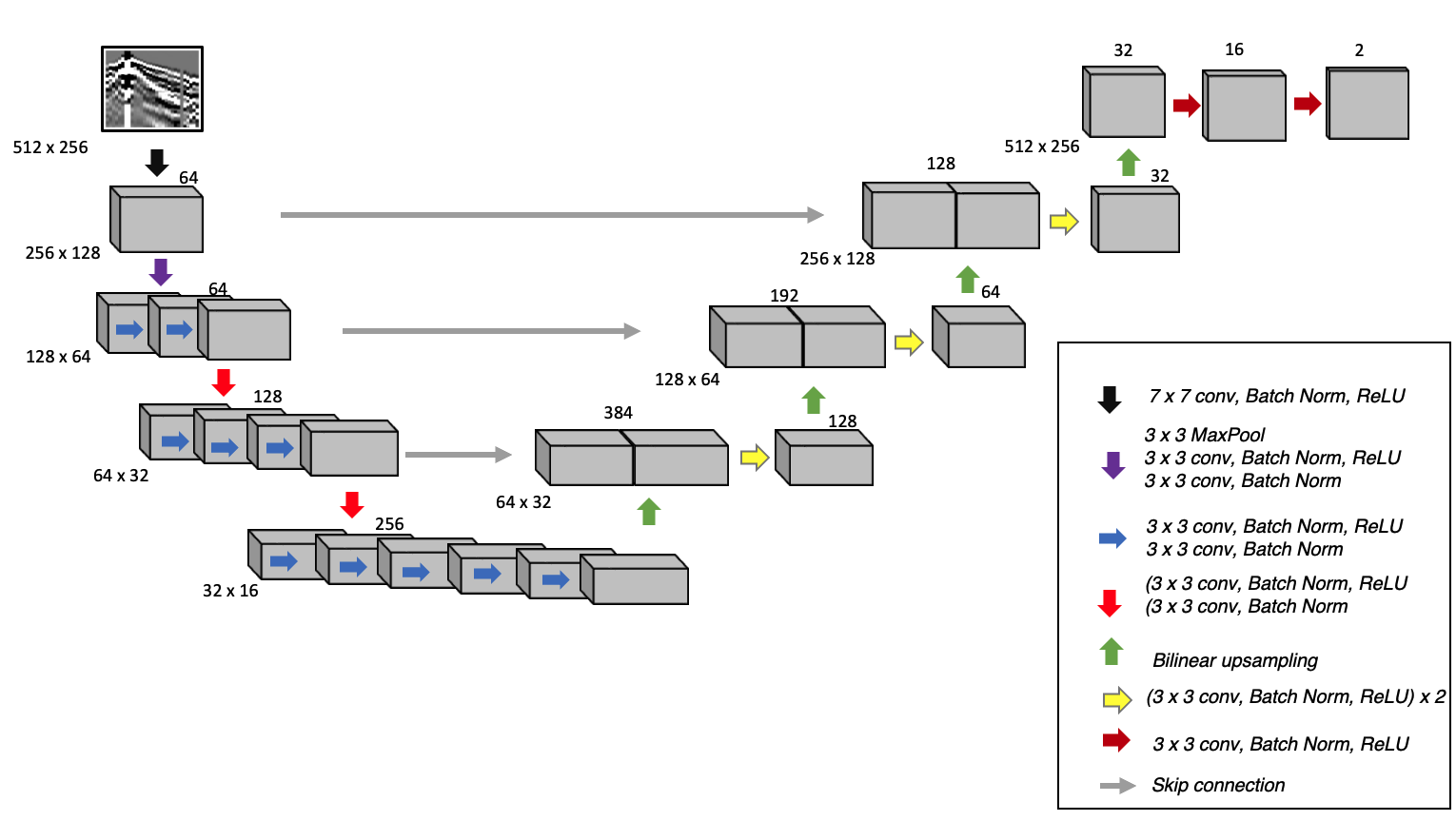}
\caption{U-Net architecture with ResNet encoder where the purple, blue, and light red arrows in the encoder show the residual blocks.
{\label{fig:architeture}}
}
\end{center}
\end{figure}

The neural network employed in this study is based on U-Net \citep{ronneberger2015u} as presented in Figure \ref{fig:architeture}.
U-Net is a fully convolutional encoder-decoder network with long skip-connections between encoder and decoder that helps to have a smoother loss leading to faster training \citep{li2018visualizing}.
The input of the network is a one-channel patch of upsampled subimages $x \in \mathbb{R}^{512 \times 256 \times 1}$. 
This input size allows to have the desired depth for the employed network. 
Through the network, we successively downsample the spatial resolution of the input to catch the higher-level features.
We use residual blocks \citep{he2016deep} to build the encoder which can helps to prevent the gradient vanishing problem \citep{zwartjes2022first}.

At the lowest level of the network, the output of the encoder is a patch of $x \in \mathbb{R}^{32 \times 16 \times 256}$.
The decoder acts as an expanding path which builds a segmentation map from the encoded features.
The decoder includes four blocks.
Each block duplicates the spatial resolution by performing a bilinear upsampling.
Using skip connection, the result of upsampling is concatenated with the output of its symmetric block in the encoder.
At each step, the concatenated tensor is passed into two convolution operators in addition to batch normalization and rectified linear unit (ReLU) operators to halve the number of channels with respect to the output of the previous decoding block.
At the end, the output of the network is a probability image $P(x) \in \left[0, 1\right]^{512 \times 256 \times 2}$ which contains two channels corresponding to the predicted probability of each class (before FB and after FB). 

\subsection*{Loss function and optimization}
In this study, the cross-entropy loss function is used to measure the accuracy of the segmentation.
This loss function for one sample (subimage) can be written as
\begin{equation}
E = -\sum_{i} y_{i}log\left(\hat{y}_{i}\right),
\label{eq:base_loss_fn}
\end{equation}
where $y_{i}$ and $\hat{y}_{i}$ are the given and predicted segments for pixel $i \in \Omega$ with $\Omega \subset \mathcal{Z}^2$.

The output of the network is the probability of each pixel to fall in one of the two segments.
The final segmentation can be obtained by passing this probability tensor into an argmax operator which can be defined as equation \ref{eq:argmax} for function $f$,
\begin{equation}
\text{argmax} f := \{x  : f(s) \leq f(x) \text{ for all } s \in X\}.
\label{eq:argmax}
\end{equation}
Equation \ref{eq:argmax} allows to build the segmentation map, $\Phi(x) \in \left[0, 1\right]^{512 \times 256 \times 1}$, from the probability image, $P(x)$.

The cross-entropy loss function (equation \ref{eq:base_loss_fn}) can be used to perform an accurate segmentation for first-break picking \citep{wu2019semiautomatic,zhu2019phasenet,maEtAl2020,zwartjes2022first}.
While equation \ref{eq:base_loss_fn} is used to train and evaluate the network for segmentation, we also used the root-mean-squared error (RMSE) to measure the error for predicting the FBs in a dataset as,
\begin{equation}
\text{RMSE} = \sqrt{\frac{\sum_{i=1}^N\left(x_i - \hat{x}_i\right)^2}{N}},
\label{eq:rms_fb}
\end{equation}
where $x_{i}$ and $\hat{x}_{i}$ are the given and predicted FB for trace $i$ in a shot with $N$ traces.
It should be mentioned that equation \ref{eq:rms_fb} is not involved in training and it is only employed to measure the accuracy of the network for first-break picking.

Employing the Adam optimization method \citep{kingma2014adam}, the network parameters are optimized to reduce the loss based on equation \ref{eq:base_loss_fn}.
In all case studies, we use a learning rate of $1\times 10^{-4}$ and each batch of training data includes 15 pairs of subimages and their corresponding labels.

\subsection*{Parameter initialization}
To train a network, loss is decreased through an iterative method.
Thereby, initial weights are required to start the training and the performance of most algorithms are affected by the initial weights of the network \citep{goodfellow2016deep}.
Even with a very large training dataset, the effects of an efficient initialization are considerable \citep{glorot2010understanding} and an ineffective initialization can prevent the training to converge \citep{goodfellow2016deep}.
 Besides, preparing a large and comprehensive dataset for training is expensive and time-consuming. 
 Transfer learning is a technique that allows us to mitigate these problems. 
 Using this technique, we can transfer the knowledge that a network has learned in one setting to improve generalization in another setting \citep{goodfellow2016deep}. 
Thereby, a network can be initialized using pretrained weights of another network and fine-tuned to pick FBs accurately.

In this study, we assess the effects of two different initializations strategies to improve the accuracy of automatic first-break picking.
In addition to random initialization, we use the weights of a pretrained resnet-34 to initialize the encoder presented in Figure \ref{fig:architeture}.
The pretrained resnet-34 is trained to solve the ImageNet classification problem \citep{deng2009imagenet}.
The ImageNet is a database of more than 14 million images and in the next section, we assess the effects of initializing the encoder using the parameters of the pretrained resnet-34. 
For this purpose, we have performed three tests and have trained three networks for each test:
\begin{itemize}
\item training using ImageNet weights with 20 epochs (ImageNet-20),
\item training using random initialization with 20 epochs (Random-20),
\item training using random initialization with 100 epochs (Random-100).
\end{itemize}
All networks are trained using Adam optimization with a learning rate of $1\times 10^{-4}$.

\section*{Results}
In this study, we use datasets from three different land surveys to perform three experiments (named Experiment 1 - 3).
Dataset characteristics of each experiment are presented in Table \ref{tbl:datasets}.
There are 244 and 291 seismic shots in Dataset 1 and Dataset 2, respectively.
Seismic source in Dataset 1 is an Isotta energizer while seismic data in Dataset 2 are obtained using a sledge hammer.

In Experiment 1, we use 21 seismic shots from Dataset 1 (less that $9\%$ of the shots) to train and evaluate the networks.
Then, the trained networks are used to perform the first-break picking on the remaining 223 seismic shots.
This experiment is conducted to study the efficiency of different initialization strategies for analyzing clean seismic data (Dataset 1).

\begin{table}[ht!]
\centering
\begin{tabular}{lccccc} 
 \hline
 Name & \makecell{Total number of \\ shots} & \makecell{Number of shots\\ for training and testing} & \makecell{Number of traces\\ per shot}  & \makecell{Number of subimages\\ per shot} & \makecell{Number of subimages\\  for training}\\ [0.5ex] 
 \hline\hline
Dataset 1     &         244 &	   21  &   101 to 201  &  4 to 8  & 133\\
Dataset 2     &         291 &      20  &   48  			&  3  	  & 54  \\
\makecell[l]{Dataset 3 \\(general)}    &         585 &      91 &   48 to 201  &  2 to 8    & 277\\
 \hline
\end{tabular}
\caption{The used datasets in this study and their characteristics.}
\label{tbl:datasets}
\end{table}

\begin{figure}[ht]
\begin{center}
\includegraphics[width=1.0\textwidth]{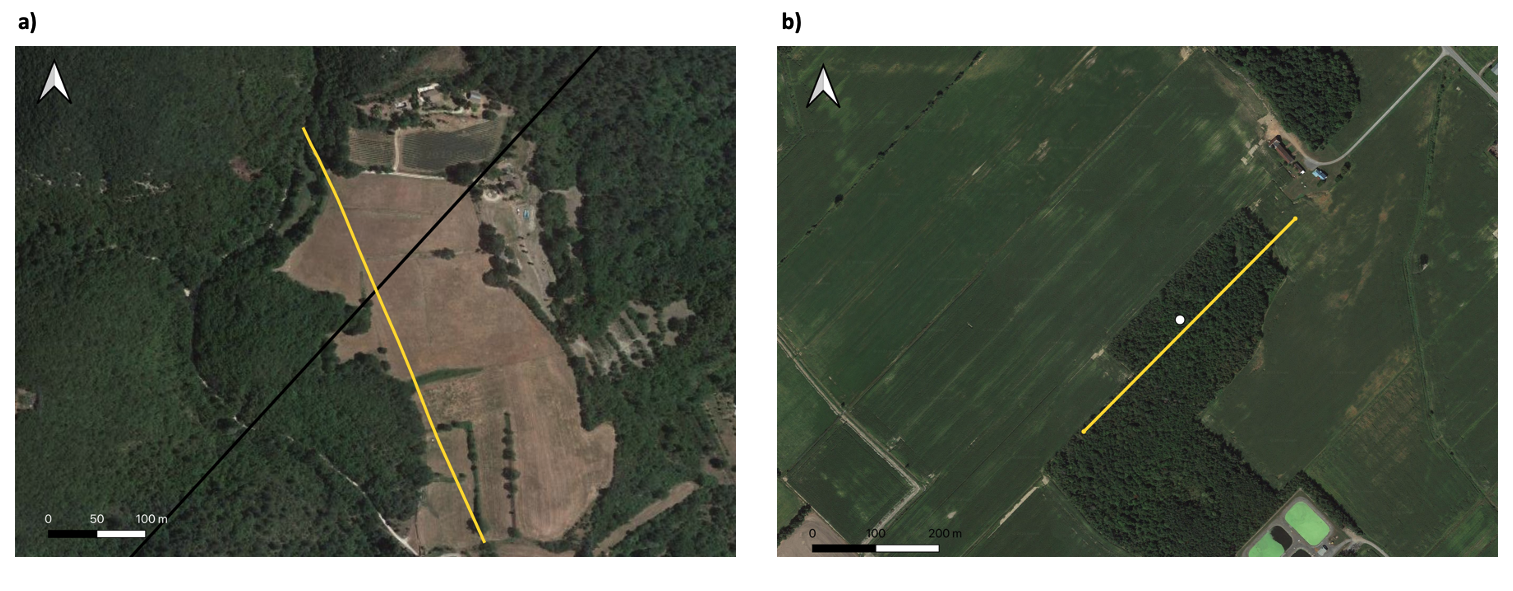}
\caption{Acquisition lines for the data studied in (a) Experiment 1 and (b) Experiment 2. Yellow lines show the acquisition profiles and black line shows the Marsanne fault in the study area. White circle denotes an exploration well in the study area.
{\label{fig:map}}
}
\end{center}
\end{figure}

The seismic data in Dataset 1 were acquired in the eastern bank of the Rhône river in France (Figure \ref{fig:map}a). 
The profile (yellow line) has a South-East to North-West direction, on the top of Cuijanet syncline crossing the Marsanne fault (black line).
In the northwestern part, the Barremian limestone is almost outcropping.
In its southeastern part, the fluvial alluvium covers the Albian and Cenomanian conglomerates, themselves covering Albian sandstone.
For this acquisition, the seismic sources were placed on the profile with a spacing of 2 m.

In Experiment 2, we analyze the network for performing first-break picking on noisy data (Dataset 2).
Only 20 seismic shots out of 291 ($\sim 7\%$) are labeled manually and used to train the network in this experiment.
Dataset 2 is highly noisy and first-break picking is very challenging even for a human operator.

This dataset is from a near-surface seismic survey acquired with a 48-channel system in Saint-Liboire in Quebec, Canada (Figure \ref{fig:map}b).
In this study, the spacing between seismic sources is 2.5 m and the receivers are placed at every 5 m.
In this project, seismic refraction tomography was used in combination with electrical resistivity tomography for characterizing the aquifer with the ultimate goal of finding an appropriate location for a water well.
To increase the SNR, seismic shots in Dataset 2 are a stack of 3 to 5 recordings at each location. 
Nevertheless, these data have a poor SNR.
Hence, this dataset is a good candidate to analyze the efficiency of the proposed automatic FB picking method with noisy data.

After studying the efficiency of different initialization strategies for picking FBs on clean and noisy seismic data, we focus on creating a general dataset (Dataset 3) through Experiment 3.
This can help us to understand the efficiency of the DL-based first-break picking if we augment the training dataset by combining the data obtained under different circumstances (i.e. various equipment, acquisition parameters, and geology of the study areas).
Dataset 3 is built by combining Dataset 1, Dataset 2, and 50 seismic shots (with picked FB) from a third project where sledge hammer is used as seismic source.
The data in these three datasets are acquired by different companies and different equipment at different locations.
It is better to have a single specialist to prepare the training dataset to reduce the subjectivity of the results.
However, the first breaks in Dataset 3 were picked by three different experts in the present case.
Although this is not ideal, it affects all three networks equally and it should not impact the results of this study, because we compare all the employed networks under the same circumstances and data preparation is the same for all these networks.

As presented in Figure \ref{fig:data_preparing} and Table \ref{tbl:datasets}, the seismic shots are converted to the different numbers of subimages (based on the number of available traces) for each project.
For each experiment, we use $90\%$ of the labeled data to train the network and the other $10\%$ as a testing set.

\subsection*{Data with high signal-to-noise ratio (Experiment 1)}
\begin{figure}[t]
\begin{center}
\includegraphics[width=1.0\textwidth]{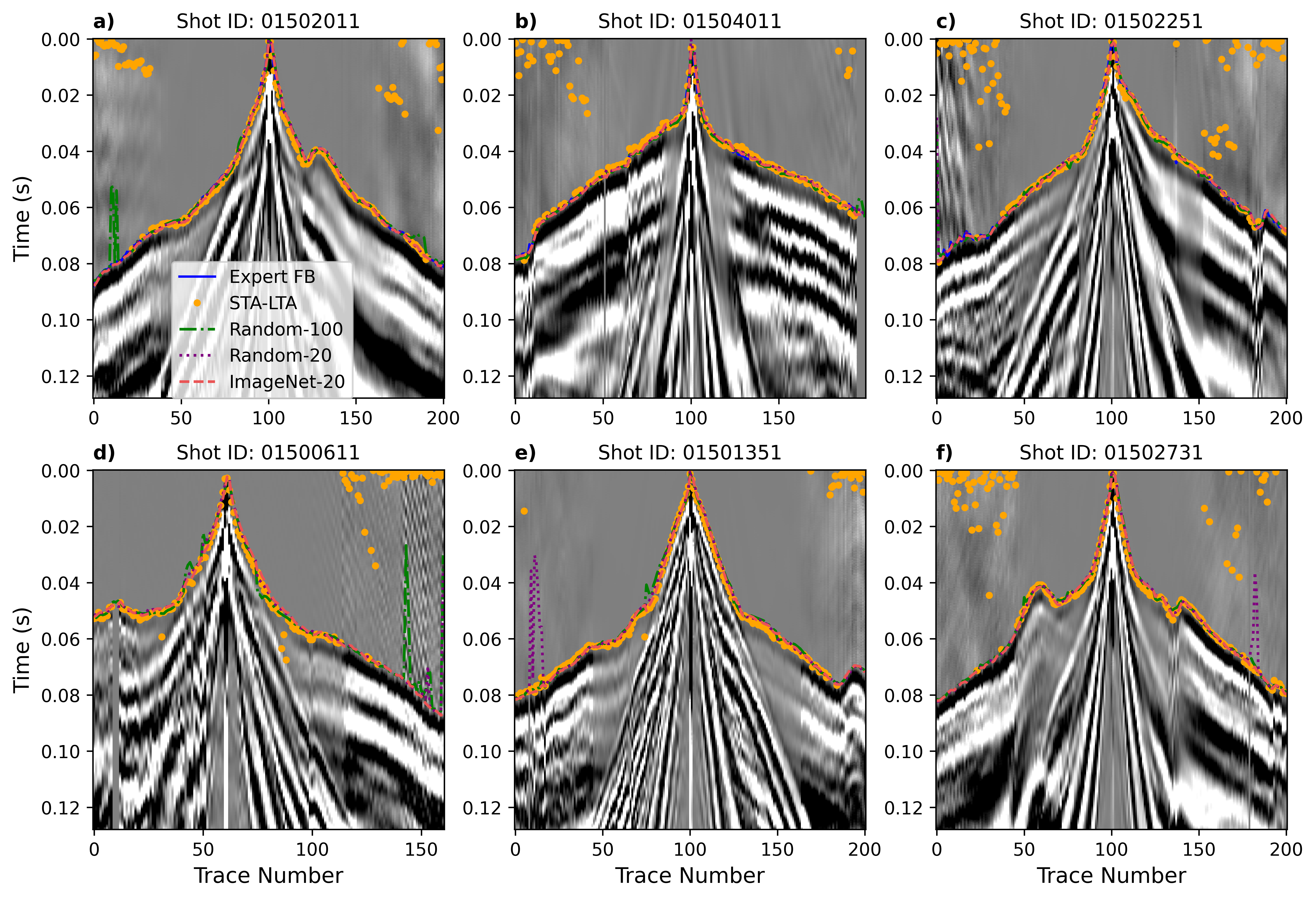}
\caption{Results of FB picking on (a-c) labeled and (d-f) unlabeled datasets for the data with high SNR. The solid blue line shows the manually picked FB while the results of STA-LTA are shown with orange dots. The dash-dotted green line shows the results using random initialization with 100 epochs. The dotted purple and the dashed red lines show the picked FB using the same number of epochs with different initialization (random and ImageNet respectively). 
{\label{fig:edf_results}}
}
\end{center}
\end{figure}

For Experiment 1, a dataset with good SNR (Dataset 1) has been used.
The training dataset for this project includes 21 seismic shots with variable number of traces per shot.
These seismic shots are divided into subimages of 30 traces with $15\%$ overlap. 
Figure \ref{fig:edf_results}a-\ref{fig:edf_results}c shows three examples of seismic shots that have been used to train the network. 
The manually picked FBs (solid blue line) are presented in addition to the FBs picked using DL and STA-LTA.
As is shown, with the same number of epochs, the network that is initialized using ImageNet (dashed red line) is more accurate than the network that is initialized using random weights (dotted purple line).
The final training losses for Random-100, Random-20, and ImageNet-20 are presented in Table \ref{tbl:segmentation_fb_loss}.
Using 100 epochs, the network with random initialization leads to the same accuracy as ImageNet-20 for segmentation of the training dataset.
However, more epochs causes overfitting as the test loss of Random-100 is higher than that of ImageNet-20 (Table \ref{tbl:segmentation_fb_loss}).
In Figure \ref{fig:edf_results}, we can also see the effects of noise on the STA-LTA results (orange dots). 
In traces with larger offset, the efficiency of STA-LTA decreases, so does the efficiency of Random-20 and Random-100.

\begin{table}[ht]
\centering
\begin{tabular}{lccc|ccc|ccc} 
 \hline
{}  &  \multicolumn{3}{c}{Experiment 1 (High SNR)} & \multicolumn{3}{c}{Experiment 2 (Low SNR)}  & \multicolumn{3}{c}{Experiment 3 (General data)} \\ [0.5ex]
{}  & Training & Test  & FB RMSE & Training & Test & FB RMSE & Training & Test & FB RMSE\\
 \hline\hline
ImageNet-20 &	   \textbf{0.026}  &   \textbf{0.039}  & \textbf{0.425} 	&  0.073 & \textbf{0.126}  & \textbf{0.941} & 0.018   &  \textbf{0.030} & \textbf{0.394}\\
Random-20  	&      0.029   & 0.072  	&   0.614 	  						&  0.070 & 0.156  	&0.990	 				&  0.020  &  0.051  & 1.0 \\
Random-100  &     \textbf{0.026}     &  0.067   	& 1.0  				&  \textbf{0.068} & 0.159  & 1.0 			&  \textbf{0.017}  &  0.052 &  0.969\\
 \hline
\end{tabular}
\caption{The segmentation loss and normalized first-break picking loss for different tests and networks.}
\label{tbl:segmentation_fb_loss}
\end{table}

In the next step, the trained networks are employed to detect the FBs in the unlabeled data.
The results are shown in Figure \ref{fig:edf_results}d-\ref{fig:edf_results}f. 
As for the test set, these shot gathers have not been seen before by the networks.
However, in this case, no first breaks were manually picked.  
Initializing the network using ImageNet leads to the most accurate result, while increasing number of epochs for training randomly initialized networks does not improve the results significantly (Table \ref{tbl:segmentation_fb_loss}).


To train the networks, the cross-entropy loss function (equation \ref{eq:base_loss_fn}) is used to evaluate the accuracy of the results, numerically.
Although the loss value shows the accuracy of each network for segmentation, this parameter is not really representative for first-break picking quality.
This is due to the fact that bad segmentation of even one pixel before the FB, can lead to erroneous picked FB (Figure \ref{fig:challenges}c).
On the other hand, bad-segmentation after the FB, does not affect the accuracy of picked FB while it still affects the value of the segmentation loss.
The final accuracy of segmentation and the normalized first-break picking errors (equation \ref{eq:rms_fb}) for Random-100, Random-20, and ImageNet-20 are presented in Table \ref{tbl:segmentation_fb_loss}.
This experiment shows that initializing a network with weights of a pretrained network leads to more accurate first-break picking for seismic data with high SNR.
%


\subsection*{Data with low signal-to-noise ratio (Experiment 2)}
For the second experiment, we focus on picking FBs in seismic data with low SNR (Dataset 2).
We have picked the FBs on 20 seismic shots (out of 291) used 18 as training data and 2 as test data. 
We divided each shot into three subimages.
Each subimage is created using 22 traces from the corresponding seismic shot with $15\%$ overlap with other subimages.

\begin{figure}[hb]
\begin{center}
\includegraphics[width=1.0\textwidth]{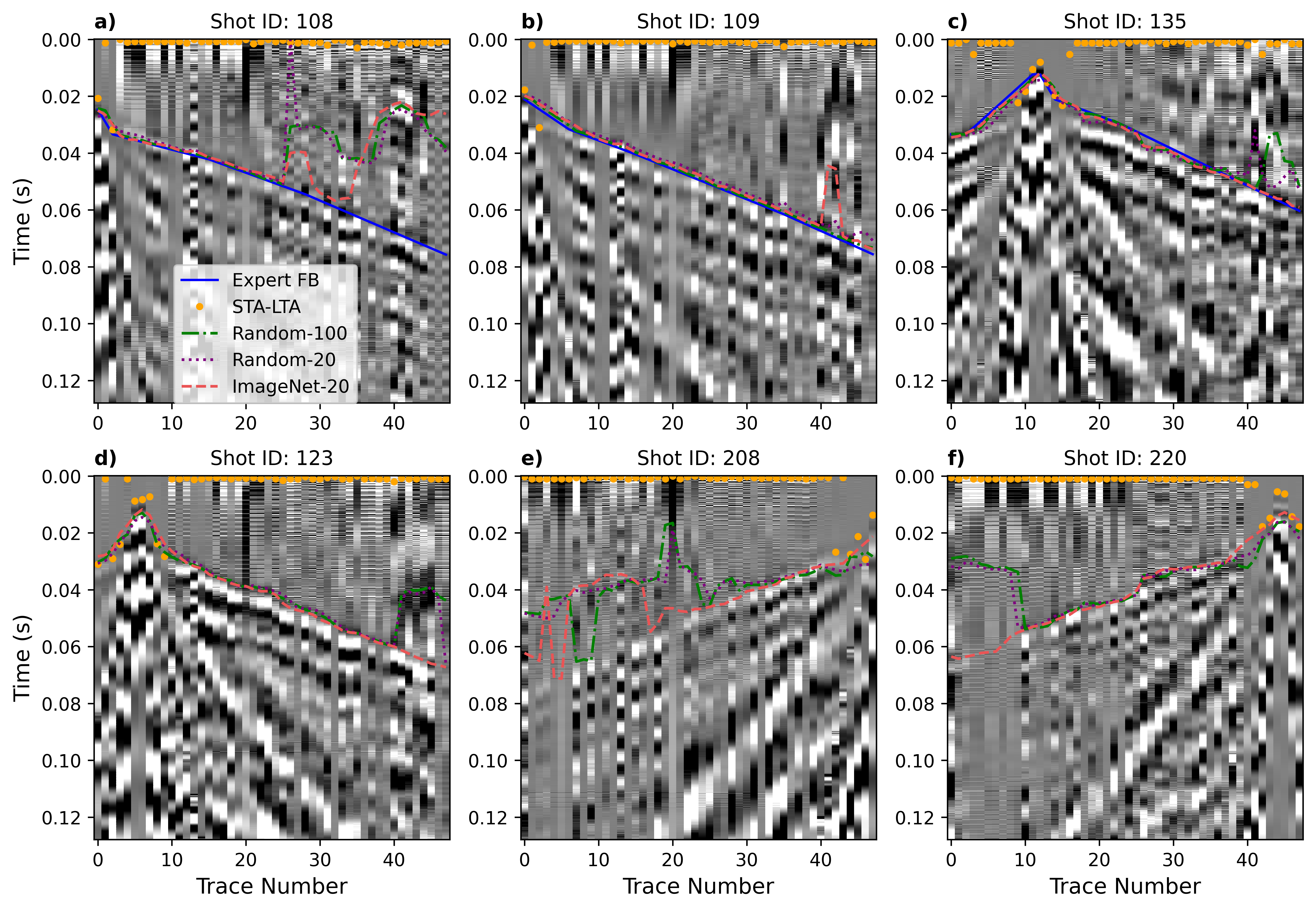}
\caption{Results of FB picking on (a-c) labeled and (d-f) unlabeled datasets for the data with low SNR. The solid blue line shows the manually picked FB while the results of STA-LTA are shown with orange dots. The dash-dotted green line shows the results using random initialization with 100 epochs. The dotted purple and the dashed red lines show the picked FB using the same number of epochs with different initialization (random and ImageNet respectively). 
{\label{fig:jeremy_results}}
}
\end{center}
\end{figure}

Figure \ref{fig:jeremy_results}a-\ref{fig:jeremy_results}c shows the results of automatic first-break picking for three examples from the training dataset.
While the performance of all three networks is comparable in most seismic shots, random initialization of the network with 100 epochs leads to higher training accuracy.
Results of STA-LTA for this dataset show that this method is completely unreliable to pick the FBs on such a noisy dataset.

The training and test losses for Experiment 2 are presented in Table \ref{tbl:segmentation_fb_loss}.
The final training losses for Random-100, Random-20, and ImageNet-20 are respectively $0.068$, $0.070$, and $0.073$. 
While Random-100 provides the highest training accuracy, the goal is to train a network that can have the highest test accuracy.
This is the case for ImageNet-20 where the network is pre-trained.
Based on the test loss, ImageNet-20 is more powerful for performing the segmentation of test dataset, with a loss of $0.126$ in comparison to Random-20 and Random-100 with losses of $0.156$ and $0.159$.
The first-break picking errors for the networks also demonstrates higher accuracy of ImageNet-20.


Applying the trained networks on the unlabeled data, we can again see the advantage of using ImageNet-20 over Random-20 and Random-100.
Using pretrained weights, the network is better at picking FBs.
Nevertheless, these data are highly noisy and even the ImageNet-20 has problem in some seismic shots (e.g. Figure \ref{fig:jeremy_results}a in training set and Figure \ref{fig:jeremy_results}e from unlabeled dataset).
It should be pointed out that picking the right FBs in these seismic shots can be also a very challenging task for an expert.
For data with this level of noise, supervision of an expert is unavoidable. 
However, for the sake of making a comparison between the studied networks, we can see that ImageNet-20 is more accurate for first-break picking and segmenting the test dataset.


\subsection*{General dataset (Experiment 3)}
For the third experiment, we mixed the labeled datasets of the first two case studies with another 50 seismic shots that have acquisition geometry similar to the seismic shots in Dataset 2.
As shown in Table \ref{tbl:datasets}, Dataset 3 includes 244 seismic shots from Dataset 1 and 291 seismic shots from Dataset 2.
The 50 seismic shots from the other study have 48 traces, and 30 traces are used to create the corresponding subimages.
This allows us to add 100 examples to the training and test datasets.
In total, the labeled dataset for this experiment includes 308 shots from which $90\%$ are used as training set (277 subimages) and $10\%$ as test set (31 subimages).

The segmentation and FB picking losses for the three networks are presented in Table \ref{tbl:segmentation_fb_loss}.
The test loss shows the better generalizability of ImageNet-20 with loss of $0.030$ in comparison with $0.051$ and $0.052$ for Random-20 and Random-100.
The first-break picking error of this experiment also shows the superiority of the ImageNet-20. 

\begin{figure}[t]
\begin{center}
\includegraphics[width=1.0\textwidth]{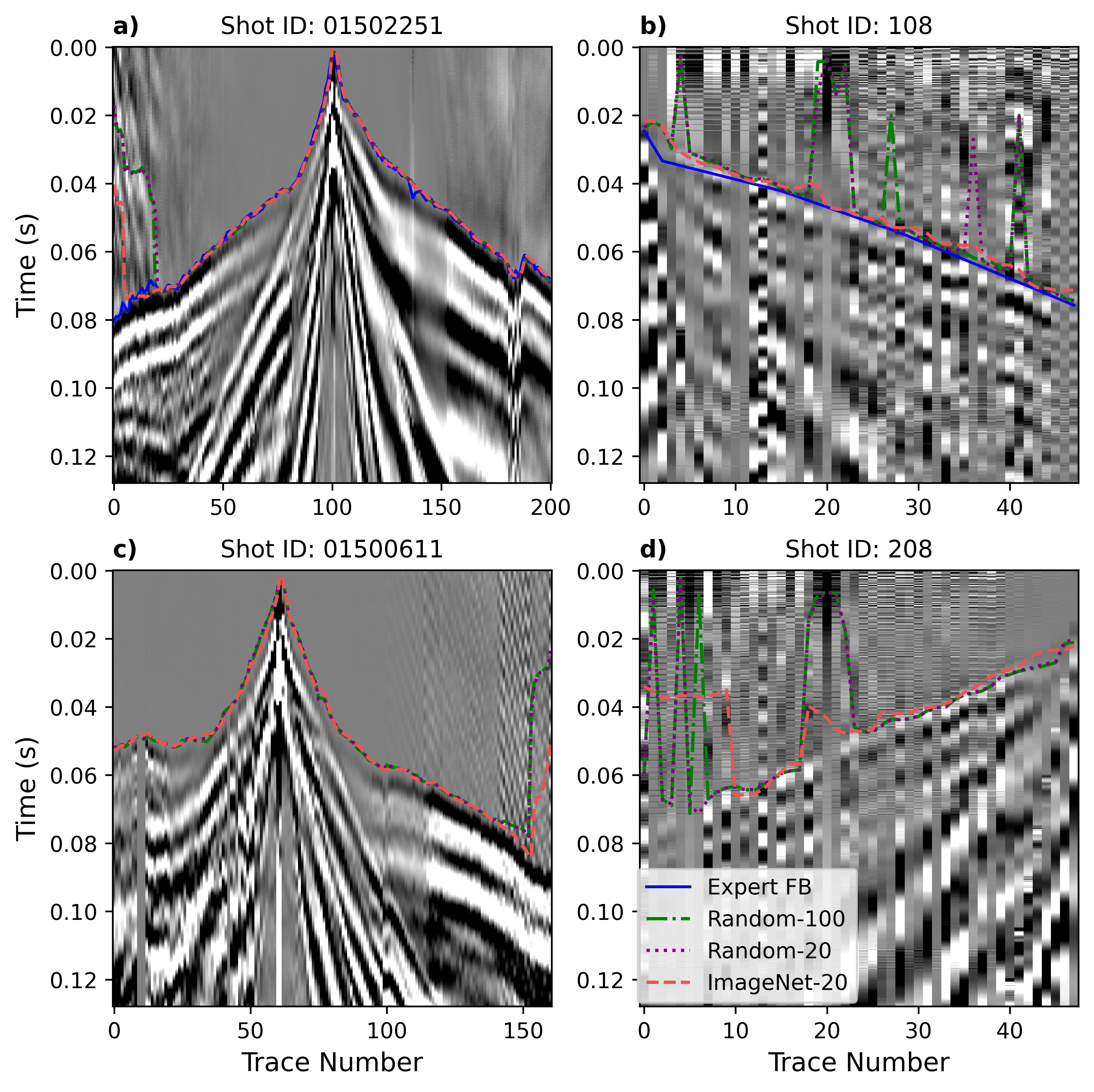}
\caption{Results for training the network on data from different projects. The solid blue line shows the manually picked FB while the dotted purple and the dashed red lines show the picked FB using the same network with different initialization (random and ImageNet respectively). The dash-dotted green line shows the result of using random initialization with 100 epochs. These examples are (a and b) training (c and d) unlabeled samples of (a and c) Dataset 1 and (b and d) Dataset 2.
{\label{fig:three_data_results}}
}
\end{center}
\end{figure}

The results of applying the trained networks on four examples are shown in Figure \ref{fig:three_data_results}.
Figure \ref{fig:three_data_results}a and \ref{fig:three_data_results}b belong to the training dataset while Figure \ref{fig:three_data_results}c and \ref{fig:three_data_results}d are part of the unlabeled dataset.
Comparing Figure \ref{fig:three_data_results}a with Figure \ref{fig:edf_results}c shows that the strategy of adding training data from different projects cannot guarantee better accuracy.
This is shown in Table \ref{tbl:loss_source} where project-based training dataset led to higher accuracy for Dataset 1 in comparison to using the general dataset.
On the contrary, Table \ref{tbl:loss_source} shows that the general dataset increased the accuracy for first-break picking of shots in Dataset 2 (comparing Figure \ref{fig:three_data_results}b with Figure \ref{fig:jeremy_results}a). 
This can be due to the fact that in addition to 20 seismic shots that are common between Dataset 2 and Dataset 3, another 50 shots have similar seismic source and acquisition geometry to Dataset 2 (Table \ref{tbl:datasets}). 
It means 78$\%$ of the shots in Dataset 3 have similar acquisition geometry and seismic source.
Thereby, training a network using this general dataset leads to higher accuracy for seismic shots in Dataset 2 while less accuracy for seismic shots in Dataset 1.

\begin{table}[ht]
\centering
\begin{tabular}{lcc|cc} 
 \hline
{}  &  \multicolumn{2}{c}{Shots from Dataset 1} & \multicolumn{2}{c}{Shots from Dataset 2} \\ [0.5ex]
{}  & Project-based Dataset & General Dataset  & Project-based Dataset & General Dataset \\
 \hline\hline
ImageNet-20   &  \textbf{0.0011} & 0.0014 &	   0.0014  &   \textbf{0.0002} \\
Random-20  	&  \textbf{0.0016} & 0.0030  &      0.0015  &   \textbf{0.0013}   \\
Random-100  &  \textbf{0.0026} & 0.0029  &    0.0015  &   \textbf{0.0013}\\
 \hline
\end{tabular}
\caption{The first-break prediction loss based on equation \ref{eq:rms_fb} for comparing the accuracy of results in case of using project-based or general datasets.}
\label{tbl:loss_source}
\end{table}

Figures \ref{fig:three_data_results}c and \ref{fig:three_data_results}d show two seismic shots from Dataset 1 and Dataset 2, respectively.
The FBs are picked using the networks that are trained with general dataset.
The results can be compared with Figures \ref{fig:edf_results}d and \ref{fig:jeremy_results}e, respectively where project-based datasets are employed to train the networks.

\clearpage
\section*{Refraction traveltime tomography}
In near surface studies, FBs are usually used to estimate a $P$-wave velocity ($V_P$) model of the subsurface.
In this section, we inverted the picked FBs of Experiment 1 and Experiment 2 to compare the estimated velocities.
This inversion is done using PyGIMLi \citep{rucker2017pygimli} and based on the shortest path algorithm \citep{moser1991shortest}.
It allows us to analyze the accuracy of picked FBs in terms of the final outcome of a near surface seismic survey: a $P$-wave velocity model.
Our assessment criterion is that the inversion result of the output of an efficient automatic FB picking method should have the minimum discrepancy with the inversion result obtained from manually picked FBs.

\begin{figure}[hb]
\begin{center}
\includegraphics[width=1.0\textwidth]{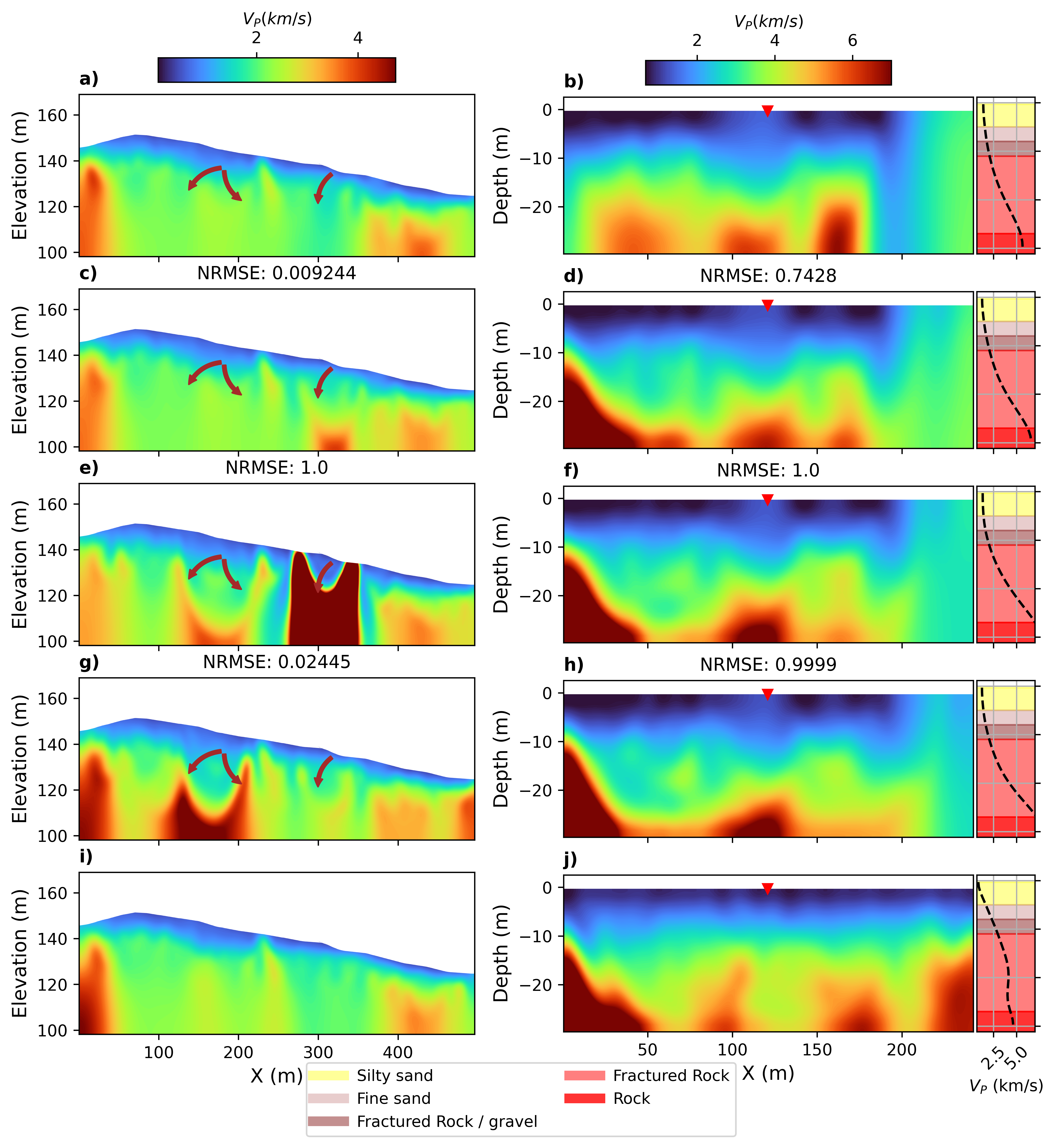}
\caption{Results of traveltime inversion for Experiment 1 (left column) and Experiment 2 (right column). Results of inversion for first breaks that are involved in training procedure picked by (a and b) an expert, (c and d) ImageNet-20, (e and f) Random-20, and (g and h) Random-100. (i and j) Estimated $V_P$ using FBs of all seismic shots picked by ImageNet-20.
 Red arrows show the main differences between the estimates from automatically picked FBs and manually-picked ones.
 Red triangle shows the well location (white circle in Figure \ref{fig:map}b). The 1D plots show the estimated velocity at the well location.
{\label{fig:inversion_results}}
}
\end{center}
\end{figure}

Estimates are shown in Figure \ref{fig:inversion_results}.
The first column shows the results for Experiment 1 and the second column presents the results obtained from Experiment 2.
Figure \ref{fig:inversion_results}a - \ref{fig:inversion_results}h denotes the estimated velocity from the picks in the labeled dataset (training and test set).
$V_P$ estimated from manually picked FBs are shown in
Figure \ref{fig:inversion_results}a and \ref{fig:inversion_results}b. 
As indicated by normalized root-mean-squared error (NRMSE) of velocity, the estimates obtained from FBs picked by ImageNet-20 (Figure \ref{fig:inversion_results}c - \ref{fig:inversion_results}d) have smaller discrepancy with the ones obtained from manually picked FBs in comparison to estimates from Random-20 (Figure \ref{fig:inversion_results}e - \ref{fig:inversion_results}f) and Random-100 (Figure \ref{fig:inversion_results}g - \ref{fig:inversion_results}h). 
Employing Random-20 and Random-100 to detect FBs in Experiment 1 leads to some non-geological structures in the velocity model that are due to inaccurate picks. 
These anomalies are shown with red arrows.

Based on our workflow, the trained network is used to obtain FBs in all seismic shots.
We then verify the obtained FBs and modify them in case of inaccuracy. 
Using more information and better ray path coverage decrease the uncertainty in traveltime inversion and leads to higher accuracy. 
Hence, as ImageNet-20 has provided the most accurate FBs (Table \ref{tbl:segmentation_fb_loss}), we have used this network to predict the FBs in all seismic shots in Dataset 1 and Dataset 2.
These FBs are then verified and inverted to obtain velocity models.
The velocity models are shown in Figure \ref{fig:inversion_results}i and \ref{fig:inversion_results}j.
These models show meaningful images of the subsurface. 
While Figure \ref{fig:inversion_results}i does not have anomalies such as the ones denoted with red arrows in other estimates of Experiment 1, Figure \ref{fig:inversion_results}j shows using more FBs can improve the accuracy of estimate.
Picking more FB in Experiment 2 led to a better model of the deeper part of our study area.
This can also be seen with comparing the 1D estimations close to the well location (red triangle) which is shown in the right side of the second column (Figure \ref{fig:inversion_results}).
The lithology of the well is color coded based of the drilling report and it can be seen that utilizing more FBs allows better constraining the inversion and leads to higher accuracy in the deeper parts.

\clearpage


\section*{Discussion}
The effects of weight initialization for DL-based automatic first-break picking is analyzed in this study.
Considering a common network and hyperparameters in three different scenarios, it is shown that using pretrained weights can improve the accuracy of picked FBs.
To make a fair comparison, the network with random initialization is trained also with more epochs.
Considering first-break picking as a segmentation problem, the network with pretrained weights leads to more accurate results for test set while it can lead to a lower training accuracy compared to the randomly initialized networks.
This shows that this network is better generalizable.
Table \ref{tbl:segmentation_fb_loss} presents the segmentation error for the three tests conducted in this study.
For these tests, Random-100 provides the lowest training loss while ImageNet-20 provides the highest accuracy for test set.
This shows that using random initialization and higher number of epochs can cause overfitting, especially with small training sets as employed in this study.
Although the ImageNet challenge is an object recognition problem, the weights of the trained network can be employed in a segmentation problem and lead to a lower test loss than using random values for initializing a network.


As the goal of this study is first-break picking, Table \ref{tbl:segmentation_fb_loss} shows the loss between manually picked FBs and predicted FBs using the different networks.
As is shown, the ImageNet-20 provides the highest accuracy and the networks with random initialization provide the lowest accuracy.

This study shows that a general training dataset can decrease the accuracy of the network in comparison with project-based training dataset if the dataset is not large enough to represent appropriately different situations such as various types of sources, noise level, geology of the study area (depth of the bedrock), and other  acquisition parameters.
A better understanding of this fact can allow to create multiple general training datasets for a specific situation (e.g. a general dataset for acquisitions with a sledgehammer as the source or acquisitions with similar geometry).

Besides, it is important to conduct more studies to analyze the effects of data preparation parameters such as the width of subimages and overlap between subimages on the efficiency of DL-based automatic first-break picking.
Although this will not affect the fact that initializing the network with pretrained weights leads to more accurate results, it will help one to develop a more effective workflow for automatic first-break picking.

The method we propose is readily applicable on field data. A workflow based on NN fine tuning comprises the following steps:
\begin{enumerate}
\item pick the first breaks of a small fraction of shots (between 5 to 10$\%$),
\item prepare the training and testing set as described in the section "data preparation",
\item fine-tune the network on the training set, and monitor overfitting with the test set,
\item apply the fine-tuned network on the remaining shots.
\end{enumerate}

This workflow can lead to a significant reduction in the time required for first-break picking. 
Indeed, the data preparation can be fully automated and is a rather straightforward task.
The network training is also fast and cheap, as the training set can be small using fine-tuning. 
For instance, training in this study was conducted on a laptop with an Apple M1 Max chip and it took 86.76 seconds and 26.74 seconds for Experiment 1 and Experiment 2, respectively.

\section*{Conclusions}
An efficient automatic first-break (FB) picking based on artificial intelligence relies significantly on the size of training dataset whose preparation and processing are costly and time consuming. 
This study focuses on reducing the size of required dataset for training a network while not compromising the accuracy. 
In this regard, we have proposed performing automatic first-break picking by fine-tuning a neural-network pretrainined on ImageNet. 
A U-Net architecture with residual encoder has been used for this study. 
Using random initialization, we have shown that the network has lower accuracy than in the case of using pretrained weights (ImageNet challenge in this study).
This strategy allows to use only 5 to 10$\%$ of seismic data from one project to train a network and perform automatic picking with high accuracy on the remaining shots.
We have shown that using first breaks obtained by random initialization can lead to non-geological features in the velocity model estimated using the FBs.
Comparing with short-term-average over long-term-average method, we have also shown that the proposed deep learning based autopicking provides more accurate results especially in the presence of noise and in seismic shots with low signal-to-noise ratio.
Finally, we have demonstrated that using a general training dataset for various projects might cause a decrease in the accuracy of picked FBs if the dataset is biased toward a specific acquisition geometry. 

\section*{Acknowledgement}
This work was supported by Mitacs through the Mitacs Elevate Program.

\section*{Data and code availability statement}
Python package for performing first-break picking based on the presented workflow is available at \url{https://github.com/geo-stack/first_break_picking}.
Data associated with this research are confidential and cannot be released.

\section*{Conflict of interest}
All authors declare that they have no known financial interests or personal relationships that could have influenced the work presented in this paper.

\bibliographystyle{seg}  
\bibliography{./fine_tuning_for_fb.bib}

\end{document}